\documentclass[%
 aip,
 amsmath,amssymb,
 reprint,%
]{revtex4-1}

\usepackage{graphicx}% Include figure files
\usepackage{dcolumn}% Align table columns on decimal point
\usepackage{bm}% bold math
\usepackage{comment}
\usepackage[utf8]{inputenc}

\usepackage{xcolor}
\usepackage{hyperref}    % must come *after* most other packages

\hypersetup{
  colorlinks   = true,
  linkcolor    = blue,
  citecolor    = purple,
  urlcolor     = cyan,
}

\makeatletter
\def\@email#1#2{%
 \endgroup
 \patchcmd{\titleblock@produce}
  {\frontmatter@RRAPformat}
  {\frontmatter@RRAPformat{\produce@RRAP{*#1\href{mailto:#2}{#2}}}\frontmatter@RRAPformat}
  {}{}
}%
\makeatother
\begin{document}

\preprint{AIP/123-QED}

\title{Effects of Unequal Electron-Ion Plasma Beta on Pressure-Strain Interaction in Turbulent Plasmas}
\author{M. Hasan Barbhuiya}
\affiliation{Department of Physics and Astronomy, Clemson University, Clemson, SC 29634}
\email{mbarbhu@clemson.edu}

\author{Subash Adhikari}
\affiliation{Department of Physics and Astronomy, University of Delaware, Newark, DE 19711}

\date{\today}% It is always \today, today,

\begin{abstract}
A common occurrence in weakly collisional space plasmas is the unequal electron-ion temperatures. The pressure–strain interaction provides a mechanism-agnostic pathway for increasing plasma internal energy through spatiotemporally local isotropic compression and volume-preserving deformation, yet its behavior under thermal disequilibrium is largely unexplored. We investigate this using five fully kinetic two-dimensional particle-in-cell simulations of undriven decaying turbulence by varying the initial electron-to-ion temperature ratios. 
By analyzing the species' internal energy density alongside a decomposition of the pressure–strain term, with a focus on the volume-preserving deformation that contains normal and shear contributions, we quantify how the initial temperature imbalance modifies the channels through which turbulence increases each species' internal energy density. The cumulative pressure-strain interaction tracks the change in internal energy for both electrons and ions, with the total deformation channel of energy conversion dominating. We discover that local changes to electron internal energy density are governed primarily by the shear deformation power density, concentrated in electron-scale current sheets, while the ion shear and normal deformation components cancel, yielding a much smaller net deformation power density that peaks around, rather than within, those electron-scale current structures.
By varying the initial temperature ratio, we find that the amplitudes and localization of deformation change, but preserve these qualitative trends.
Together, these results show how thermal disequilibrium could shape species-dependent turbulent ``heating rate", measured via pressure-strain interaction and approximated via only its shear deformation part, and provide a framework for interpreting energy evolution and conversion in space plasmas where unequal species temperature is the norm.
\end{abstract}

\maketitle

\section{Introduction} 
\label{sec:intro}
%\textbf{Big picture paragraphs of Why and who cares?}
Astrophysical and heliospheric plasmas are, in general, weakly collisional and often turbulent in nature \cite{retino_observation_2007,marsch2006_LivingReview,quataert_2003}.
In such systems, plasma species can and do exist with unequal plasma betas (defined as the ratio of gas pressure and magnetic pressure), since strong collisions that drive systems to a late-time state where species temperatures reach equality are absent. An outcome is a significant departure from local thermodynamic equilibrium (LTE) may exist in turbulent plasmas, and it is not well understood how irreversible conversion of energy stored as bulk flow energy of the plasma species and electromagnetic (EM) field energy into thermal energy, \textit{i.e.,} dissipation takes place at the smallest scales of driven and/or decaying turbulence \cite{Howes17,Matthaeus20,Howes_2024_JPP}. 
As borrowed from the collision-dominated neutral fluid theory \cite{batchelor67}, this conversion of energy is irreversible, and it thermalizes plasmas, \textit{i.e.,} phase space densities become more Maxwellian as all available energy is dissipated. However, the conversion of bulk and EM energy into the internal energy of weakly collisional plasma remains formally reversible, in the Vlasov-Maxwell formalism \cite{Yang17,Cassak_FirstLaw_2023} without collisions driving dissipation.

To properly model and study weakly collisional plasma systems, by capturing the key physics of non-LTE plasmas, it thus becomes imperative to use kinetic models. Gyrokinetic models, which by construction are reduced models \cite{hallatschek2004thermodynamic,Howes06}, based on the guiding center approximation \cite{northrop1961guiding}, rely on small deviations from the equilibrium Maxwellian phase space densities. Past studies consistently found that plasma beta fundamentally controls the partitioning of available energy between ions and electrons, which increases their internal energy, and has been called ``heating rates" in the literature, a phrase derived from collision-dominated neutral fluid theory. 
The quantitative relationships between $\beta$ and the ratio of the ``heating rate" of ions and electrons, however, vary.
Within the limit of gyrokinetics, through comprehensive beta scans \cite{Kawazura_2019_PNAS}, it has been discovered that the ion-to-electron heating rate ratio $Q_i/Q_e$ is an increasing function of $\beta_i$ and that ``heating rate" ratio is insensitive to the ion-to-electron temperature ratio $T_i/T_e$, indicating that beta rather than temperature ratio is the primary control parameter. 
Moreover, in the relativistic regime, where ions are sub-relativistic, and electrons are ultra-relativistic, it has been found that ions preferentially gained energy, gaining up to an order of magnitude more than electrons, with the energy partition described by a simple empirical formula based on the ratio of electron-to-ion gyroradii \cite{Zhdankin_2019_PRL}. We should note, however, that this includes both degrees of particle kinetic energy, \textit{i.e.,} internal and bulk flow. Given the relevance to high-energy astrophysical systems such as hot accretion flows onto black holes, the bulk flow energy may likely be a higher fraction of the total energy gain. 

The mechanisms through which plasma species gain internal energy have been shown to vary, as well, in the gyrokinetic limit.
At low  $\beta_i$ ($\le 1$), nonlinear phase mixing was found to dominate gain in ion internal energy, while at high $\beta_i$ ($\gg 1$), linear phase mixing (Landau damping) becomes the dominant mechanism \cite{Kawazura_2019_PNAS}; due to the choice of the model used, electron heating mechanisms were not explored. Later studies \cite{Howes_2024_JPP,Adkins_2025_ApJ} have discovered that for low plasma beta, $\beta_i \le 0.5$, electron Landau and transit-time damping dominate, while above this value, ion Landau damping and transit-time damping become significant, thereby increasing the ion internal energy. Another transition occurs at $\beta_i = 10$ for what is called ``kinetic viscous heating", mediated by temperature anisotropies \cite{Arzamasskiy_2023_PRX}, which could be linked to pressure-strain interaction \cite{batchelor67} that contains anisotropic pressure terms through the volume preserving deformation term called $\mathrm{Pi-D}$ \cite{Yang17,yang_PRE_2017,Cassak_PiD1_2022,del_sarto_pressure_2018}. 

In fully kinetic simulations, modeled using particle-in-cell (PIC) simulations, which allow plasmas to be arbitrarily far from LTE, a more limited beta range ($\beta_i = \beta_e$ = 0.3, 0.6, 1.2) study\cite{parashar_2018_ApJL} observed that increasing plasma beta leads to greater total ``heating", with ions gaining internal energy preferentially more than electrons. The average ``heating rate" of ions increased with beta, while it decreased for electrons, leading to a higher internal energy ratio over time. This study also identified $\beta_e = \beta_i > 1$ as a critical value where pressure effects overcome magnetic tension effects, leading to less orderly coherent structures and increased ion heating. The key observation of increased ``viscous"-like energization at higher beta values matches the observations made in the gyrokinetic limit \cite{Howes_2024_JPP}. %Roy \textit{et al.}  
A phenomenological model based on both simulations and spacecraft data was later developed \cite{Roy_2022_ApJ}, where the beta dependence was included through a relationship found through fitting of 2.5D and 3D PIC simulations and spacecraft data rather than from a first-principle approach. Multiple candidate mechanisms, including stochastic heating, cyclotron resonances, kinetic instabilities, magnetic pumping, and reconnection, were identified as potential heating processes, with a high focus on the pressure-strain interaction. 

The preferential ion ``heating" at high beta has direct implications for understanding how the temperature ratio of species evolves in heliophysical plasmas. For near-Earth solar wind and in low-beta magnetosheath turbulent plasmas, with $\beta \sim1$, large amplitude driving could explain why ions are hotter than electrons \cite{parashar_2018_ApJL}. In astrophysical plasmas, there could be implications for low-luminosity accretion flows around black holes, where the ``heating ratio" can affect accretion rates and outflow dynamics \cite{Zhdankin_2019_PRL}.

As hinted earlier, the ``heating" mechanism of kinetic ``viscous" heating could be the pressure–strain interaction term, which describes the local change in internal energy density due to strain rate effects, due to its coupling with both isotropic and anisotropic parts of the pressure tensor.
The three different physical mechanisms, in an orthogonal Cartesian system, captured by pressure-strain interaction are pressure dilatation, describing LTE compressive changes to the internal energy density, and how normal and shear deformations change the internal energy density, respectively \cite{Cassak_PiD1_2022,Barbhuiya_PiD3_2022}.
Within the gyrokinetic framework, the challenge of numerical modeling due to the multiscale nature of plasma turbulence has been highlighted, which, though necessitating the use of reduced models, comes at the expense of losing key kinetic physics \cite{Howes_2024_JPP}. 
Within the fully kinetic framework that uses PIC simulations, the need for further examination of parameter space, including different electron and proton betas, has been noted \cite{parashar_2018_ApJL}.  
Heating ratios obtained using spacecraft and simulation data required fitting, and identified the need for further study on compressive heating, locally and at intermediate scales \cite{Roy_2022_ApJ} [which we address in a separate study \cite{Adhikari_2026_PRE}].
In the present work, we use fully kinetic decaying (undriven) turbulence simulations to study how unequal initial electron–ion temperature ratios affect the local change in the species' internal energy density due to pressure-strain interaction and its shear and normal deformation parts to evaluate their separate roles. 
Our results indicate that the deformation channel dominates the reversible conversion of flow energy into internal energy for both electrons and ions in periodic simulations, with shear deformation providing the largest contribution. We also discover that spatiotemporally, the shear deformation dominates normal deformation for electrons, and is moderately anti-correlated, while for ions, the anti-correlation is much stronger, leading to muted ``total" deformation. Furthermore, we find that inside the strong current regions, which have been suspected of localized regions producing most ``heating" \cite{Parashar_2016_ApJ}, shear and normal deformations are anti-correlated, but their collective effect on changing the ion internal energy is weak, suggesting that most of the change is happening not at electron-scale structures, which is true for electrons instead.

This paper is organized as follows: Section~\ref{sec:theory} discusses the theoretical background, while in Section~\ref{sec:sim} we provide the details of the simulations used. Section~\ref{sec:results} gives the results from the PIC simulations, and finally, in Section~\ref{sec:conclusions}, we summarize our results, discuss the implications, and present the future outlook.

\section{A Review of the Key theoretical concepts\label{sec:theory}}  
For fully-ionized weakly collisional plasmas, the time evolution %equation for
of internal energy density ${\cal E}_{{\rm int},\sigma}$ for species $\sigma$ is given by
\begin{equation}
    \frac{\partial {\cal E}_{{int},\sigma}}{\partial t} + \boldsymbol{\nabla} \cdot ({\cal E}_{{int},\sigma} {\bf u}_\sigma + {\bf q}_\sigma)  =  - ({\bf P}_\sigma \cdot \boldsymbol{\nabla}) \cdot {\bf u}_\sigma, \label{eq:intengevolve}
\end{equation}
where  ${\cal E}_{{int},\sigma} = (3/2) n_\sigma k_B {\cal T}_\sigma$,  $n_\sigma = \int d^3v f_\sigma$ is the number density with $f_\sigma$ as the phase space density, $k_B$ is Boltzmann's constant, and ${\cal T}_\sigma = (2/3k_Bn_\sigma)\int d^3v (1/2) m_\sigma v_\sigma^{\prime 2} f_\sigma$ is the effective temperature, ${\bf v}$ as the velocity space coordinate, ${\bf v}_\sigma^\prime = {\bf v} - {\bf u}_\sigma$ is the peculiar velocity and ${\bf u}_\sigma = (1/n_\sigma) \int d^3v {\bf v} f_\sigma$ is the bulk flow velocity. ${\bf q}_\sigma = \int d^3v (1/2) m_\sigma v_\sigma^{\prime 2} {\bf v}_\sigma^\prime f_\sigma$ gives the vector heat flux density, and ${\bf P}_\sigma$ is the pressure tensor with elements $P_{\sigma,jk} = \int d^3v m_\sigma v_{\sigma,j}^\prime v_{\sigma,k}^\prime f_\sigma$. 

We presented Eq.~\ref{eq:intengevolve} in the conservative form, %of a conservation law, 
where the spatiotemporally local internal energy density can change in the Eulerian frame due to non-zero divergence of flux associated with it and heat flux density, and due to the term that couples the pressure tensor with strain rate tensor, giving pressure-strain interaction, which acts as a source/sink term. Pressure-strain interaction describes the rate of change of internal energy density resulting interaction between the bulk flow strain effects and pressure, which contains LTE and non-LTE elements, and thus describes both LTE and non-LTE pathways to internal energy density change. The conservative form of Eq.~\ref{eq:intengevolve} demonstrates that one key property for systems that are isolated, closed, infinite, or periodic (as is the case in simulations considered in the present study). It shows that the change in the ``global" evolution of internal energy, \textit{i.e.,} spatial-volume integrated form of Eq.~\ref{eq:intengevolve}, which in turbulence literature, \textit{e.g.,} \cite{Roy_2022_ApJ} called the ``heating rate", is only due to volume-integrated pressure-strain interaction, for each species $\sigma$ separately.
The pressure-strain interaction has, thus been, utilized as a proxy for ``heating/cooling" rate in recent studies, both theoretical/numerical \cite{Hazeltine_2013_PoP,Sitnov_2018_GRL,parashar_2018_ApJL,Du_2018_ApJ,yang_scale_2019,Pezzi19,song_forcebalance_2020,Du_energy_2020_PRE,Fadanelli21,Arro22,Yang_2022_ApJ,Cassak_PiD1_2022,Barbhuiya_PiD3_2022, Barbhuiya_PRE_2024, Barbhuiya_phases_2025,Hellinger_2022_ApJ} and satellite observations \cite{Chasapis_2018_ApJ,Zhong_2019_GRL,Bandyopadhyay_2020_PRL,bandyopadhyay_energy_2021,zhou_measurements_2021,Wang_2019_GRL,Burch_PoP_2023,Roy_2022_ApJ} in weakly collisional space plasmas.

In the recent literature, the pressure-strain interaction is written in a number of equivalent ways.  $-({\bf P}_\sigma \cdot \boldsymbol{\nabla}) \cdot {\bf u}_\sigma = - {\bf P}_\sigma : \boldsymbol{\nabla} {\bf u}_\sigma = -P_{\sigma,jk} (\partial u_{\sigma,k} / \partial r_{\sigma,j})$.  
Using $\boldsymbol{\nabla} {\bf u}_\sigma = {\bf S}_\sigma + \boldsymbol{\Omega}_\sigma$, where ${\bf S}_\sigma$ is the symmetric strain rate tensor and $\boldsymbol{\Omega}_\sigma$ is the anti-symmetric strain rate tensor, it is found that ${\bf P}_\sigma : \boldsymbol{\Omega}_\sigma = 0$ since ${\bf P}_\sigma$ is a symmetric tensor; the physical consequence of this property is that rigid body rotations do not contribute to pressure-strain interaction \cite{del_sarto_pressure_2016,Cassak_PiD1_2022} and that pressure-strain interaction only has contributions from the pure straining motion of the fluids.

We suppress the subscript $\sigma$ henceforth and use it only to allay any confusion in our discussions. 
By coupling the decomposition of the pressure tensor into its isotropic ${\cal P}$ and non-isotropic part $\boldsymbol{\Pi}$, called the deviatoric pressure tensor, and the decomposition of strain rate tensor \cite{batchelor67,del_sarto_pressure_2016,Yang17,del_sarto_pressure_2018} into ${\bf S} = (1/3) {\bf I} (\boldsymbol{\nabla} \cdot {\bf u}) + \boldsymbol{{\cal D}}$, pressure-strain interaction can be decomposed into
\begin{equation}
  -{\bf P}:\mathbf{S} = - {\cal P} (\boldsymbol{\nabla} \cdot {\bf u}) - \Pi_{jk} {\cal
    D}_{jk}. \label{eq:pdelu}
\end{equation}
Here, ${\bf I}$ is the identity tensor and $\delta_{jk}$ is the Kronecker delta. $\boldsymbol{{\cal D}}$ is the traceless strain rate tensor with elements given by ${\cal D}_{jk} = (1/2) \left( {\partial u_{j}}/{\partial r_k} + {\partial u_{k}}/{\partial r_j} \right) - ({1}/{3}) \delta_{jk} (\boldsymbol{\nabla} \cdot {\bf u})$.

The first term (including the minus sign), called pressure dilatation, describes the power density associated with changing the internal energy density due to bulk flow compression/expansion, as this term is associated with $(1/3) {\bf I} (\nabla \cdot {\bf u})$ that describes volume changing compressive/expansive effects. 
The second term (including the minus sign), called \cite{Yang17} ${\rm Pi-D}$, captures the power density associated with changing the internal energy density due to %incompressive 
volume-preserving deformations \cite{del_sarto_pressure_2018}, which are volume-preserving deformations as encapsulated by $\boldsymbol{{\cal D}}$.
${\rm Pi-D}$ was also called ``collisionless viscosity'' \cite{yang_PRE_2017} because it is analogous in form to collisional viscous heating \cite{Hazeltine_2013_PoP}, where collisional closures are present, as in the case of magnetohydrodynamics and neutral fluid hydrodynamics \cite{batchelor67}. Recent works have utilized ${\rm Pi-D}$ to calculate ``effective" collisionality in a global average sense, in weakly collisional kinetic turbulence \cite{yang_2024_MNRAS,bandyopadhyay_2023_PoP,adhikari2025estimation}.

Recently, Cassak and Barbhuiya \cite{Cassak_PiD1_2022} noted that the volume-preserving deformation effects captured by $\boldsymbol{{\cal D}}$ consist not only of normal deformation, \textit{i.e.,} due to purely normal flows, but also of shear deformation, \textit{i.e.,} due to purely sheared flow. Thus, 
$\boldsymbol{{\cal D}}$ was decomposed \cite{Cassak_PiD1_2022} into a normal deformation tensor $\boldsymbol{{\cal D}}_{{\rm normal}}$ and a shear deformation tensor $\boldsymbol{{\cal D}}_{{\rm shear}}$, so that $\boldsymbol{{\cal D}} = \boldsymbol{{\cal D}}_{{\rm normal}} + \boldsymbol{{\cal D}}_{{\rm shear}}$.
Here, ${\cal D}_{{\rm normal},jk} = [(\partial u_j / \partial r_j) - (1/3) (\nabla \cdot {\bf u})]\delta_{jk}$ (with no sum on $j$) isolates normal deformation physics, \textit{i.e.,} how flows in each direction change along only that direction.
${\cal D}_{{\rm shear},jk} = (1/2)(\partial u_j / \partial r_k+\partial u_k / \partial r_j)$ for $j \neq k$ and ${\cal D}_{{\rm shear},jj} = 0$ (no sum on $j$) isolates shear deformation physics, \textit{i.e.,} how flows in each direction change only along the other two orthogonal directions.
This led to writing ${\rm Pi-D}$ as the sum of two terms \cite{Cassak_PiD1_2022},
\begin{equation}
{\rm Pi-D} = {\rm Pi-D}_{{\rm normal}} + {\rm Pi-D}_{{\rm shear}},
\end{equation}
where ${\rm Pi-D}_{{\rm normal}} = - \boldsymbol{\Pi} : \boldsymbol{{\cal D}}_{{\rm normal}}$ and ${\rm Pi-D}_{{\rm shear}} = -\boldsymbol{\Pi}:\boldsymbol{{\cal D}}_{{\rm shear}}$. In Cartesian coordinates, these terms are 
\begin{subequations}
\begin{align}
\mathrm{Pi\!-\!D}_{\mathrm{normal}}
  &= -(\Pi_{xx}{\mathcal D}_{xx} + \Pi_{yy}{\mathcal D}_{yy} + \Pi_{zz}{\mathcal D}_{zz}) \nonumber\\
  &= -\left(\Pi_{xx}\frac{\partial u_x}{\partial x}
            +\Pi_{yy}\frac{\partial u_y}{\partial y}
            +\Pi_{zz}\frac{\partial u_z}{\partial z}\right), \label{eq:piddeform} \\
\mathrm{Pi\!-\!D}_{\mathrm{shear}}
  &= -(2\Pi_{xy}{\mathcal D}_{xy} + 2\Pi_{xz}{\mathcal D}_{xz} + 2\Pi_{yz}{\mathcal D}_{yz}) \nonumber\\
  &= -\Bigl[P_{xy}\!\left(\frac{\partial u_x}{\partial y}+\frac{\partial u_y}{\partial x}\right)
            +P_{xz}\!\left(\frac{\partial u_x}{\partial z}+\frac{\partial u_z}{\partial x}\right)\nonumber\\
  &\qquad\qquad\qquad\;+P_{yz}\!\left(\frac{\partial u_y}{\partial z}+\frac{\partial u_z}{\partial y}\right)\Bigr]. \label{eq:psincompress}
\end{align}
\end{subequations}

Consequently, the pressure-strain interaction is decomposed into three terms to isolate the power density that goes into changing the internal energy density due to the dilatation term, normal deformation term, and shear deformation term, given by \cite{Cassak_PiD1_2022}
\begin{equation}
-{\bf P}:\mathbf{S} = -{\cal P} (\boldsymbol{\nabla \cdot {\bf u}}) + {\rm Pi-D}_{{\rm normal}} + {\rm Pi-D}_{{\rm shear}}.
\end{equation}  

\section{Simulations \label{sec:sim}}
We carry out numerical simulations of decaying turbulence using the massively parallel particle-in-cell code {\tt p3d} \cite{zeiler:2002}. The simulations are fully three-dimensional in velocity space and 2.5-dimensional in physical space, so all vector quantities retain three components, while one spatial direction—the out-of-plane $\hat{z}$ direction, is taken to be invariant. The code advances particles using a relativistic Boris stepper \cite{birdsall91a} and evolves the electromagnetic fields with a trapezoidal leapfrog scheme \cite{guzdar93a}. To enforce Poisson’s equation and remove spurious electric-field divergence, {\tt p3d} applies a multigrid cleaning method \cite{Trottenberg00}. All simulations employ periodic boundary conditions in the two spatial directions ($\hat{x}$ and $\hat{y}$).

All simulation outputs are expressed in code-normalized units. Magnetic fields are normalized to a reference field strength $B_0$. Time is normalized to the inverse ion cyclotron frequency, $\Omega_{ci}^{-1} = (q_i B_0 / m_i c)^{-1}$, where $q_i$ and $m_i$ denote the ion charge and mass, and $c$ is the speed of light. Spatial scales are normalized to the ion inertial length $d_i = c/\omega_{pi0}$, with $\omega_{pi0} = (4\pi n_0 q_i^2/m_i)^{1/2}$ the ion plasma frequency based on a reference density $n_0$. Velocities are normalized to the ion Alfv\'en speed $c_{A0} = B_0/(4\pi m_i n_0)^{1/2}$, and temperatures to $m_i c_{A0}^2/k_B$. Finally, power densities are reported in units of $(B_0^2/4\pi)\Omega_{ci}$.

For numerical expedience, we employ reduced (and therefore non-physical) values of the speed of light, $c = 15$, and the electron-to-ion mass ratio, $m_e/m_i = 0.04$. Each simulation uses a square, periodic domain of size $L_x \times L_y = 37.3912 \times 37.3912$ resolved by a $1024 \times 1024$ grid, with 6,400 particles per grid (PPG). The high PPG significantly suppresses particle-in-cell noise at the small scales of interest. The initial out-of-plane magnetic field oriented along $\hat{z}$ is uniform, and has magnitude $B_0 = 1$.%, and is oriented along $\hat{z}$. 
We initialize the system with Alfv\'enic velocity and magnetic field fluctuations with random phases, excited in the wavenumber band $k \in [2,4] \times 2\pi/L_x$ with a flat spectrum. The initial root-mean-square amplitudes of both the velocity and magnetic fluctuations are set to 0.25, resulting in a very small initial cross helicity. %(less than 0.1). 
The initialization procedure follows that of a previous study with a domain four times larger~\cite{parashar_2018_ApJL}. However, we do not expect the choice of simulation domain to affect the evolution of all forms of energies, including electron and ion heating rates, based on the conclusion of an earlier study \cite{parashar_2015_ApJ}, which found a system size of close to 40 $d_i$ to be adequate to capture large-scale features. 
The characteristic nonlinear time associated with the largest turbulent structures is $\tau_{nl} = L_x/[2\pi (\delta b_{\rm rms}^2 + \delta v_{\rm rms}^2)^{1/2}] \simeq 17$, and we evolve each simulation for approximately $4\tau_{nl}$.

At initialization, the number density is uniform and set to the reference value $n_0 = 1$. Both species, electrons and ions, are initialized with drifting Maxwellian velocity distribution functions at their respective uniform temperatures, $T_e$ and $T_i$, and each species drifts with the local bulk flow velocity. The smallest characteristic time scale in these simulations is the inverse electron plasma frequency, $\omega_{pe}^{-1} \simeq 0.0133$. We choose a particle time step of $\Delta t = 0.005$, with the electromagnetic fields advanced using a smaller time step of $\Delta t/3$. 

The five simulations analyzed in the present work (that were previously used in a separate yet related study \cite{Adhikari_2026_PRE}) differ in their values of the temperature ratio $T_e/T_i$. Table~\ref{tab:simparameters} summarizes the key distinctions, including the smallest relevant kinetic length scale, either the electron Debye length $\lambda_{De}$ or ion Debye length $\lambda_{Di}$, for each temperature ratio. The grid length in both spatial directions is fixed at $\Delta = 0.03651$ for all five simulations. As indicated in Table~\ref{tab:simparameters}, for $T_e/T_i = 0.25$ and $4$ the smallest kinetic scale is not resolved; for $T_e/T_i = 0.5$ and $2$ it is marginally resolved (equal to the grid spacing); and for $T_e/T_i = 1$ it is well resolved. To satisfy Poisson’s equation and improve energy conservation, the electric field is cleaned every 40 particle time steps. As a result, the total energy remains conserved to within 0.1\% by the final simulation time of $t = 75$ for all five $T_e/T_i$ cases.

We apply a recursive smoothing procedure on the raw simulation data to mitigate the effect of PIC noise in the two-dimensional plots of quantities in the Sec.~\ref{sec:results}. We smooth over a width of four cells four times, compute spatial derivatives, and then apply the same four-cell smoothing four times again using the native IDL function SMOOTH, which employs a boxcar average. 
We evaluated different smoothing widths and numbers of iterations to determine an appropriate choice. The adopted parameters preserve sharp spatial variations and yield clearer two-dimensional structures and one-dimensional cuts through the domain while avoiding oversmoothing (not shown). This is because the net effect of this smoothing procedure is similar to employing a Gaussian filter with a size of roughly 3.5 cells, impacting data over roughly 4 cells, which still preserves $d_e$ scale electron-scale %coherent 
structures.

\begin{table}
    \centering
    \begin{tabular}{|c|c|c|c|c|}
    \hline
       $T_e/T_i$ & $T_e$  & $T_i$ & smallest length scale & $\beta_{\text {total,in-plane}}$ \\
    \hline 
        0.25 & 0.15 & 0.60 & 0.02582 ($\lambda_{De}$) & 1.5\\
        0.5  & 0.30 & 0.60 & 0.03651 ($\lambda_{De}$) & 1.8\\
        1    & 0.60 & 0.60 & 0.05164 ($\lambda_{De}~\text{\&}~\lambda_{Di}$) & 2.4\\
        2    & 0.60 & 0.30 & 0.03651 ($\lambda_{Di}$) & 1.8\\
        4    & 0.60 & 0.15 & 0.02582 ($\lambda_{Di}$) & 1.5\\
        \hline
    \end{tabular}
    \caption{The five electron ($T_e$) and ion ($T_i$) temperatures ratio cases studied in the present work, with $T_e$ and $T_i$ at initialization listed in the second and third columns. The fourth column presents the smallest length scale and whether it is the electron or the ion Debye length. The last column provides the total in-plane plasma $\beta$ (without including the out-of-plane mean field).}
    \label{tab:simparameters}
\end{table}

\section{Results\label{sec:results}}

\begin{figure}
    \centering
    \includegraphics[width=1\linewidth]{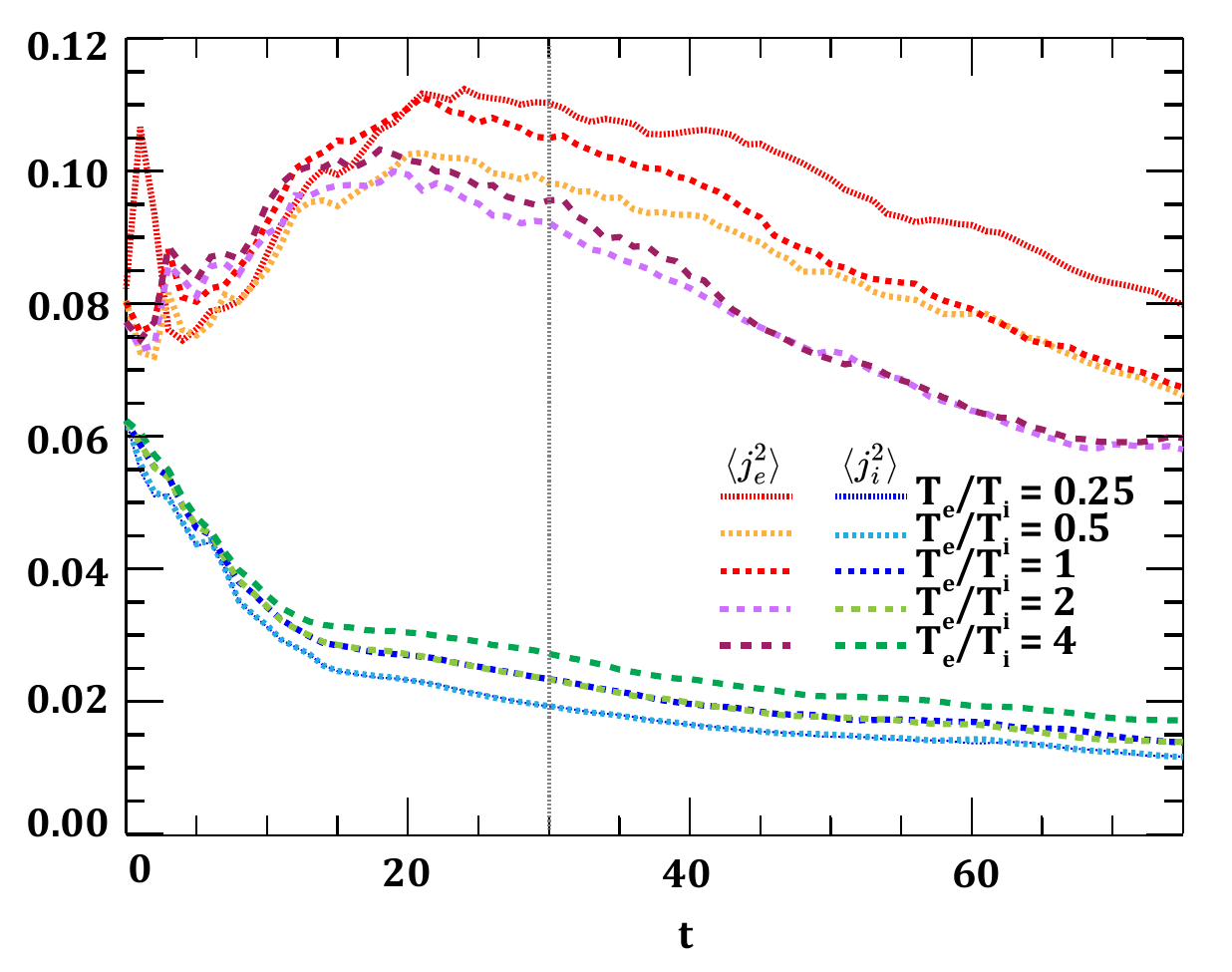}
    \caption{Mean square current for electrons in red-purple shades and ions in blue-green shades is presented as it evolves in time as a function of initial electron-to-ion temperature. The vertical gray line denotes the time chosen for comparative analysis $t=30\approx 2\tau_{nl}$.}
    \label{fig:jsquare_compare}
\end{figure}

\begin{figure*}
    \centering
    \includegraphics[width=1\linewidth]{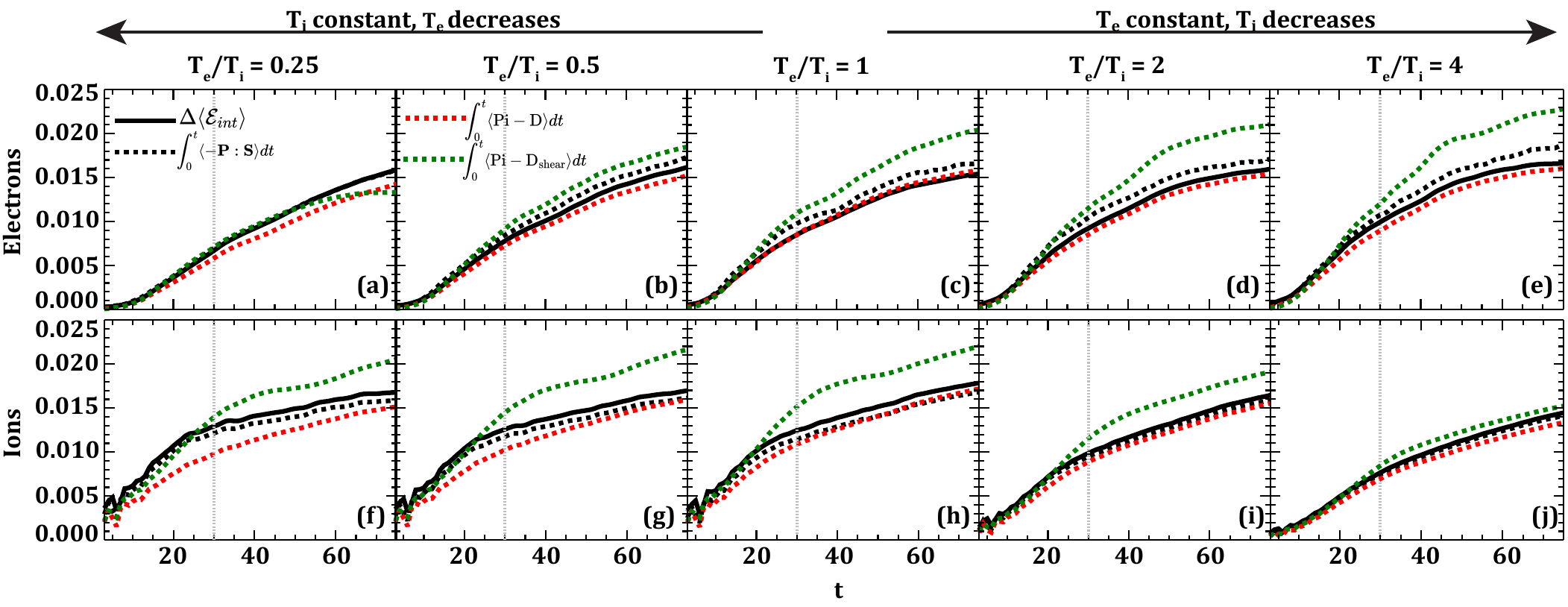}
    \caption{Change in electrons [panels (a)-(e)] and ions [panels (f)-(j)] internal energy $\Delta \langle \mathcal{E}_{int} \rangle$ (in solid black) is shown for all five simulations and is compared to the cumulative sum of system-averaged pressure-strain interaction (in black dotted), its total deformation part ${\rm Pi-D}$ (in red dotted), and its shear deformation part ${\rm Pi-D}_{\rm shear}$ (in green dotted). The gray dotted line in each panel denotes $t=30$.}
    \label{fig:Eint_PS_PiD_PiDS_compare}
\end{figure*}

We begin the discussion with an overview of the simulations via mean square current, %of the (unforced) turbulence as it decays, 
as a function of $T_e/T_i$, shown in Fig.~\ref{fig:jsquare_compare}. 
The mean square current for both species is plotted for the five simulations, with line plots in the shades of red-purple for the electron mean square current $\langle j^2_e \rangle$ and in the shades of blue-green for the ion mean square current $\langle j^2_i \rangle$. As with previous simulations \cite{parashar_2018_ApJL} that utilized similar initializations, at $t=0$, $\langle j^2_i \rangle$ is close but less than $\langle j^2_e \rangle$, with electrons becoming the dominant current carriers as time evolves in all five simulations. We find slight variations in how $\langle j^2_e \rangle$ evolves in time, as a function of $T_e/T_i$, such as the final value of $\langle j^2_e \rangle$ decreasing as the initial electron temperature is increased. The vertical gray line is at $t=30$, at which we compare all five simulations, with the caveat that certain variations in local electron and ion quantities between the five simulations will be present. 
The comparison time is approximately $2\tau_{nl}$, by which all five simulation systems have reached the maximum value of the mean-square current, as shown in Fig. 1. This is indicative of fully developed turbulence since $\langle j^2 \rangle$ can be associated with dissipation in the usual MHD sense, making a comparison of turbulent activity in these simulations appropriate.

From the form of Eq.~\ref{eq:intengevolve}, it is evident that the cumulative sum of system-averaged pressure-strain interaction $\int^t_0 \langle -\mathbf{P}:\mathbf{S}_\sigma \rangle dt$ is what provides the change in the change in the internal energy $\Delta \langle \mathcal{E}_{int,\sigma} \rangle$ of plasma species $\sigma$. Previous works \cite{Yang_2022_ApJ,yang_2024_JGR} have confirmed it in decaying turbulent systems that had a bigger simulation domain size and in reconnecting systems. 
We want to separate the contribution to $\int^t_0 \langle -\mathbf{P}:\mathbf{S}_\sigma \rangle dt$ from the normal deformation and shear deformation. 
In Fig.~\ref{fig:Eint_PS_PiD_PiDS_compare}, we plot $\Delta \langle \mathcal{E}_{int,\sigma} \rangle$ (in solid black) and find that it agrees well with $\int^t_0 \langle -\mathbf{P}:\mathbf{S}_\sigma \rangle dt$ (in black dotted) for both electron [panels (a)-(e)] and ions [panels (f)-(j)]. 
However, looking at only the deformation channel, \textit{i.e.,} $\int^t_0 \langle {\rm Pi-D}_\sigma \rangle dt$ (in red dotted), we find that though it acts as a good proxy for change in internal energy density for both species, it is slightly lower than the pressure-strain interaction for all five cases for both species, implying that though the main channel for internal energy increase is non-LTE, i.e., ${\rm Pi-D}_\sigma$, the isotropic LTE channel, \textit{i.e.}, pressure dilatation, still plays a small part in increasing internal energy. 
Furthermore, the shear deformation channel, \textit{i.e.,} $\int^t_0 \langle {\rm Pi-D}_{{\rm shear},\sigma} \rangle dt$ (in green dotted) alone is found to, in general, overestimate the increase in internal energy for both species. Looking at electrons, for the cases with $T_e$ fixed, we find that the overestimation from the shear deformation channel remains approximately the same as $T_i$ decreases [panels (c) to (e)]. But, as $T_e$ is decreased while keeping $T_i$ fixed [panels (c) to (a)], we find that the shear deformation's overestimation decreases, implying that the shear deformation effects dominate over normal deformation effects as the initial available internal energy is lowered. These trends are repeated by ions; when $T_e$ is fixed, and $T_i$ is decreased, the overestimation from the shear deformation channel decreases [panels (h) to (j)], while as $T_e$ is decreased while keeping $T_i$ fixed [panels (h) to (f)], the shear deformation's overestimation remains approximately the same. These results imply that a) normal deformation effects are anti-correlated with shear deformation effects in a global sense, and b) as the species' initial temperature decreases, the contribution from the normal deformation channel decreases significantly, making the shear deformation contribution roughly equal to the total contribution from the %incompressible 
net deformations. These two implications are true in a global sense, when the whole system is considered, so the next steps are to analyze whether it is true locally in position space. 

\begin{figure*}
    \centering
    \includegraphics[width=1\linewidth]{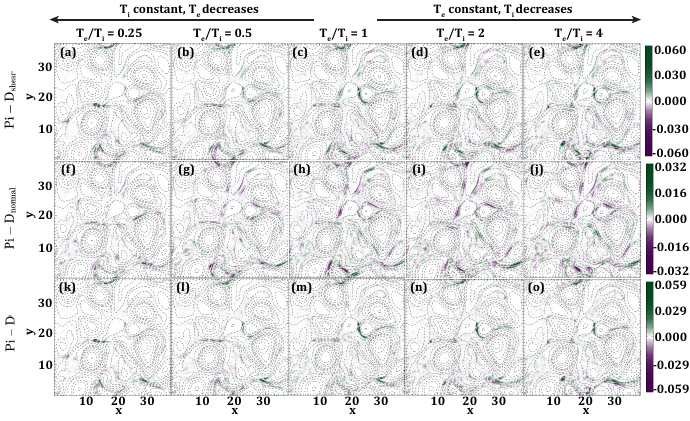}
    \caption{The shear deformation [panels (a)-(e)], normal deformation [panels (f)-(j)], and their combination, total deformation [panels (k)-(o)] for all five $T_e/T_i$ cases are shown for electrons. The color bar value of each quantity is set by the absolute maximum of the minimum and maximum seen in all five simulations at $t=30$. The in-plane magnetic fields, a proxy for interacting eddies, are shown by dashed black lines.}
    \label{fig:PiD_decomp_electrons}
\end{figure*}

\begin{figure*}
    \centering
    \includegraphics[width=1\linewidth]{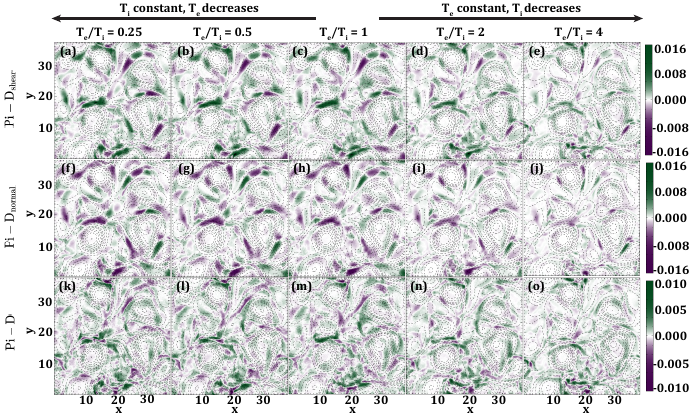}
    \caption{Figure \ref{fig:PiD_decomp_electrons} repeated for ions, for all five simulations at $t=30$.}
    \label{fig:PiD_decomp_ions}
\end{figure*}

Previous work\cite{Barbhuiya_PiD3_2022} looked at the spatiotemporally local variation in pressure-strain interaction and its shear deformation component for only electrons, \textit{i.e.,} ${\rm Pi-D}_{{\rm shear},e}$ in a symmetric reconnection simulation. Here, in Fig.~\ref{fig:PiD_decomp_electrons} and \ref{fig:PiD_decomp_ions}, we analyze both species' shear and normal deformation parts, and how they contribute to the net %incompressible 
deformation captured by ${\rm Pi-D}$. For each set of quantities at the time of analysis $t=30$, the color bar is determined by the absolute maximum of the minimum and maximum value of each quantity in all five simulations.

Fig.~\ref{fig:PiD_decomp_electrons} shows that ${\rm Pi-D}_{\mathrm{shear},e}$ [panels (a)-(e)] is concentrated within thin, electron-scale intermittent structures that are tens of $d_i$ long, between the larger ion-scale eddies (shown by the contours of in-plane magnetic field in dashed lines). 
The amplitude of ${\rm Pi-D}_{\mathrm{shear},e}$ systematically decreases as the initial $T_e$ is decreased [panels (c) to (a)], while keeping $T_i$ fixed. As the initial species temperature is decreased, so do the diagonal pressure tensor elements. Previous works \cite{Hesse04,Adhikari_2025_PoP} have shown that the off-diagonal pressure tensor elements are affected by the change in flow gradients and in-plane diagonal pressure tensor elements. The gradient scale, which for a strong guide field case is likely between $d_e$ and $\rho_e$ (electron gyroradius based on the thermal speed), at which flows change, decreases with a decrease in temperature, which by itself increases sheared flow effects. 
But, the effect of decreasing diagonal pressure tensor elements, which are larger to begin with, wins and thus ${\rm Pi-D}_{\rm shear}$ decreases, in general.  Certain dissimilarities are evident when we scan the five simulation cases at $t=30$ while the overall picture remains quasi-similar. For example, in the $T_e/T_i=0.25$ case, a pronounced quadrupolar ${\rm Pi-D}_{\mathrm{shear},e}$ feature forms around $(x,y)=(14,17)$. Development of stronger and thinner current sheet structures are expected for lower $T_e$ cases [see Fig.~\ref{fig:jsquare_compare}] where, in some electron-scale intermittent structures, the effect gradient flow changes wins due to gradient scales being smaller and sharper, thus producing a much stronger ${\rm Pi-D}_{\mathrm{shear},e}$. Similar structural differences, \textit{i.e.}, sharper gradients at the electron-scale intermittent structures when $\beta$ is varied have been seen in previous work \cite{parashar_2018_ApJL}.

Turning to the electron normal deformation shown in Fig.~\ref{fig:PiD_decomp_electrons} [panels (f)-(j)] which the reconnection studies \cite{Barbhuiya_PiD3_2022,Adhikari_2025_PoP} largely ignored, we find that it is generally weaker than ${\rm Pi-D}_{\mathrm{shear},e}$ and tends to be locally anti-correlated with it whenever the two are co-located. 
Strong flow shear regions, such as where neighboring eddies slide past one another, tend to generate shearing effects at the electron-scale regions between eddies. However, these turbulent interactions are complex and tend to also have normal flow components present, leading to anti-correlated non-zero normal flow deformation.
Although normal flow gradients ($\partial_j u_j$, with $j=x,y$) increase slightly as the characteristic scales shrink as $T_e$ is decreased, the reduction in diagonal pressure is stronger and therefore dominates, leading to an overall decrease in ${\rm Pi-D}_{\mathrm{normal},e}$ seen in panels (h) to (f). 

As expected, we do note the presence of localized variations in electron ${\rm Pi-D}_{\mathrm{shear},e}$ and ${\rm Pi-D}_{\mathrm{normal},e}$ at the time of analysis in the five simulations due to the intrinsic time dependent nature of decaying turbulence, which stems from the complicated interactions of eddies. But unsurprisingly, the overall picture in ${\rm Pi-D}_{\mathrm{shear},e}$ [panels (c) to (e)] and ${\rm Pi-D}_{\mathrm{normal},e}$ [panels (h) to (j)] with $T_e$ fixed while $T_i$ is lowered, remains unchanged. Looking at ${\rm Pi-D}_e$ in panels (k) to (o), we discover that it is mostly due to ${\rm Pi-D}_{\mathrm{shear},e}$ as also discovered in a recent reconnection study with strong guide field \cite{Adhikari_2025_PoP}. We discover that due to the anti-correlation between ${\rm Pi-D}_{\mathrm{shear},e}$ and ${\rm Pi-D}_{\mathrm{normal},e}$, the total %incompressible 
deformation effects are slightly damped in localized regions. An example of this can be seen for the case $T_e/T_i=1$, in the coherent structure centered around $(x,y)=(15,22)$, which has ${\rm Pi-D}_{\mathrm{shear},e}>0$,  but ${\rm Pi-D}_{\mathrm{normal},e}<0$, giving a weak ${\rm Pi-D}_e>0$, underscoring the importance of including the normal deformation physics in our analysis. 

We now shift our focus to ions in Fig.~\ref{fig:PiD_decomp_ions}, where panels (a)-(e) show ${\rm Pi-D}_{\mathrm{shear},i}$, panels (f)-(j) show ${\rm Pi-D}_{\mathrm{normal},i}$, and panels (k)-(o) show their sum, \textit{i.e.,} ${\rm Pi-D}_i$ for all five simulations. Before a deeper discussion of the variations in the ion flow deformation quantities, we first develop the key differences seen relative to electron deformation quantities. 
We firstly notice that both the shear and normal deformation parts for electrons are larger than ions, irrespective of the $T_e/T_i$ case. This is due to the bulk flow velocity strain rate tensor elements $\partial_j u_k$ being larger for electrons, as the gradient scale for less massive electrons is smaller and the flow speeds are higher, relative to more massive ions, which move more slowly. We also note that the amplitude of ${\rm Pi-D}_{\mathrm{shear},i}$ is close to ${\rm Pi-D}_{\mathrm{normal},i}$, unlike what was seen for electrons, where ${\rm Pi-D}_{\mathrm{shear},e} > {\rm Pi-D}_{\mathrm{normal},e}$.
Secondly, as ${\rm Pi-D}_{\mathrm{shear},i}$ is found to be more strongly anti-correlated with ${\rm Pi-D}_{\mathrm{normal},i}$ [compared to what we see for electrons] in ion-scale current sheets instead of electron-scale structures, their sum, \textit{i.e.}, ${\rm Pi-D}_i$ comes out be almost a factor of six smaller than ${\rm Pi-D}_e$ as also seen in a previous studies that employed similar simulation parameters and in other studies \cite{bandyopadhyay_energy_2021}. 
Lastly, unlike electron deformation quantities, ion deformation terms are found to be weak, but non-zero in large-scale eddies, due to demagnetized ions in the ion-scale islands producing anisotropic pressure-tensor elements, and thus producing ion deformation effects. 

Focusing on the variation in ${\rm Pi-D}_{\mathrm{shear},i}$ as a function of $T_e/T_i$ shown in Fig.~\ref{fig:PiD_decomp_ions}[panels (f)-(j)], we find that as the initial $T_i$ is decreased, while keeping $T_e$ fixed, the amplitude of ${\rm Pi-D}_{\mathrm{shear},i}$ [panels (c) to (e)] and ${\rm Pi-D}_{\mathrm{normal},i}$  [panels (h) to (f)] in the ion-scale coherent features generally decreases, similar to what is found for ${\rm Pi-D}_{\mathrm{shear},e}$, due to similar reasons discussed previously for electrons. As with electrons, localized exceptions due to the nature of decaying turbulent systems do exist, such as ion-scale coherent structure centered around $(x,y)=(15,5)$ for the $T_e/T_i=4$ case [panel (e)], where ${\rm Pi-D}_{\mathrm{shear},i}$ is seen to be stronger than what is seen elsewhere in the simulation domain, which leads to strong ${\rm Pi-D}_i$ as well [panel (o)].

\begin{figure*}
    \centering
    \includegraphics[width=1\linewidth]{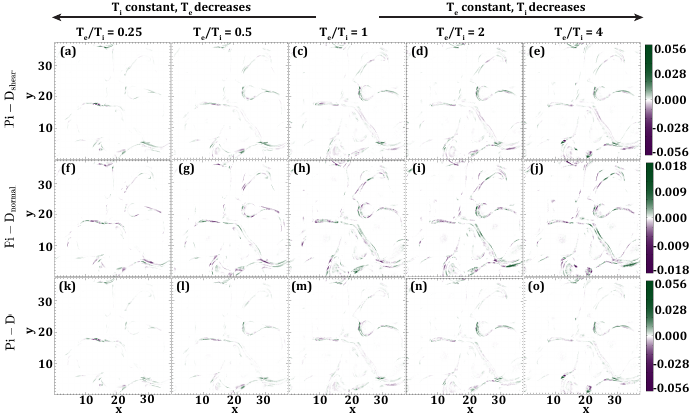}
    \caption{The shear deformation [panels (a)-(e)], normal deformation [panels (f)-(j)], and their combination, total deformation [panels (k)-(o)] for all five $T_e/T_i$ cases are shown for electrons after applying the strong current sheet conditioning to only capture regions electron shear, normal, and total deformation are non-zero in strong intermittent structures. The color bar value of each quantity is set by the absolute maximum of the minimum and maximum seen in all five simulations at $t=30$. The in-plane magnetic field lines, a proxy for interacting eddies, are removed for clarity.}
    \label{fig:PiD_decomp_electrons_conditioned}
\end{figure*}

\begin{figure*}
    \centering
    \includegraphics[width=1\linewidth]{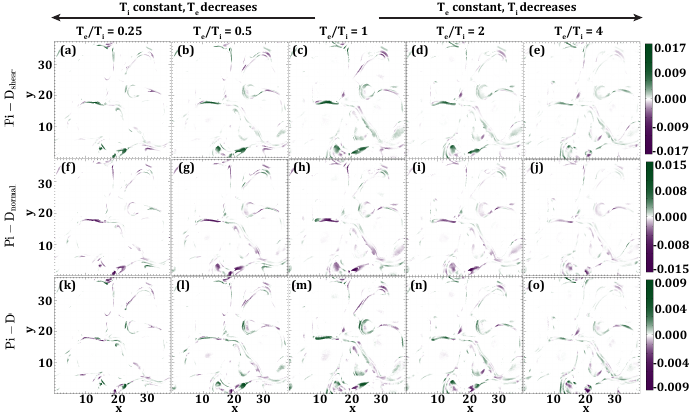}
    \caption{Figure ~\ref{fig:PiD_decomp_electrons_conditioned} is repeated for ions for all five simulations after applying the strong current sheet conditioning to data shown in Figure ~\ref{fig:PiD_decomp_ions}.}
    \label{fig:PiD_decomp_ions_conditioned}
\end{figure*}

As one goal of the present study is to find the heating rate for both species as a function of initial temperature ratios, but with a focus on strong intermittent structures instead of the whole simulation domain, we need to capture only the contribution from the strong current sheet regions for both electron and ion  ${\rm Pi-D}$ and its decompositions. To that end, we use the out-of-plane current density, which is mostly carried by electrons $J_{ez}$ (plots not shown), and condition ${\rm Pi-D}$ and its decompositions for both species. The idea is to only pick out spatial regions where ${\rm Pi-D}$ and its decompositions are non-zero while $J_{ez}$ satisfies a certain threshold. To find the threshold, we start with the probability distribution function of $J_{ez}$ at $t=30$ for each simulation, followed by finding the standard deviation. $J_{ez}$ values between the $\pm 0.5$ standard deviation are neglected, thus providing us with the threshold for strong current regions. We also tried using the full-width half maximum window to set the threshold, but found that it included large-scale eddy structures as an unintended consequence and thus was not a viable option (plots not shown).  Lastly, using the threshold $J_{ez}$ values, we create a ``filter mask" and apply it to the electron and ion ${\rm Pi-D}$ and its decompositions. The results are shown in Figs.~\ref{fig:PiD_decomp_electrons_conditioned} and \ref{fig:PiD_decomp_ions_conditioned}.

Fig.~\ref{fig:PiD_decomp_electrons_conditioned} displays the electron data, without the in-plane contours, with conditioned ${\rm Pi-D}_{\mathrm{shear},e}$ in panels (a)-(e), ${\rm Pi-D}_{\mathrm{normal},e}$ in panels (f)-(j), and ${\rm Pi-D}_e$ in panels (k)-(o). Comparing the visible structures seen between the unconditioned and conditioned deformation quantities, we see that the strong current sheet structures are well captured in the regions with the highest amplitudes of both shear and normal deformations. %${\rm Pi-D}_{\mathrm{shear},e}$. However, an analogous comparison with ${\rm Pi-D}_{\mathrm{normal},e}$ implies that regions with stronger normal deformation were ignored by the conditioning, indicating that they were likely not in strong current sheet structures, considered by the conditioning applied to the data.
Qualitatively, we also find that in the strong current sheet structures, shear deformation effects are dominantly more positive, whereas normal deformation effects are more negative, but weaker. But this weak normal deformation does locally lower the contribution of ${\rm Pi-D}_{\mathrm{shear},e}$ to the total deformation, as we see in plots of ${\rm Pi-D}_e$ [see panels (k) to (o)], along the current sheet structures that are approximately 5$d_i$ long at approximately $y=5$.
Physically, electrons demagnetize in the $d_e$-scale strong intermittent structures, producing non-LTE phase space densities that have non-zero off-diagonal pressure tensor elements, $P_{xz}$ and $P_{yz}$, given the presence of a strong mean magnetic field. The strong out-of-plane current means that the flow shear effects due to how $u_z$ changes in  $\hat{x}$ and $\hat{y}$ are the strongest, which couples to non-zero $P_{xz}$ and $P_{yz}$, thus producing strong ${\rm Pi-D}_{\mathrm{shear},e}$ [see Eq.~\ref{eq:psincompress}]. However, the normal deformation effects are due to how in-plane flows are changing in in-plane directions, which are weak to begin with. Coupled to the in-plane components of the deviatoric pressure tensor's diagonal elements, which are small as well (since $\Pi_{jj}=P_{jj}-\mathcal{P}$, where $j$ is $x,y$), produce weak ${\rm Pi-D}_{\mathrm{normal},e}$ [see Eq.~\ref{eq:piddeform}] signature in the strong current sheet regions.
[For a more in-depth kinetic interpretation of the shear and normal deformations, readers are referred to Cassak and Barbhuiya \cite{Cassak_PiD1_2022}.]
When associated with decreasing initial $T_e$ [panels (c) to (a) and (h) to (f)], it explains the overall reduction in amplitudes of conditioned shear and normal deformation components for electrons.

While examining the unconditioned ${\rm Pi-D}_{\mathrm{shear},e}$ in Fig.~\ref{fig:PiD_decomp_electrons}, we discussed one exception of localized region with higher unconditioned ${\rm Pi-D}_{\mathrm{shear},e}$ when $T_e$ is lowered with $T_i$ fixed. Similar exceptions are seen in Fig.~\ref{fig:PiD_decomp_electrons_conditioned} as well. For example, the strong current sheet structure centered around $(x,y)=(13,5)$ in panel (e) for the $T_e/T_i=4$ case highlights an unexpectedly strong bipolar structure in conditioned ${\rm Pi-D}_{\mathrm{shear},e}$, that leads to a strong bipolar conditioned ${\rm Pi-D}_e$ in panel (o). 

Focusing on ions deformation terms shown in Fig.~\ref{fig:PiD_decomp_ions_conditioned}, we first notice that decreasing $T_e$ (while keeping $T_i$ fixed) reduces the spatial extent over which conditioned ${\rm Pi-D}_{\mathrm{shear},i}$ [panels (c) to (a)],  ${\rm Pi-D}_{\mathrm{normal},i}$ [panels (h) to (f)], and ${\rm Pi-D}_i$ [panels (k) to (m)] are non-zero. Their amplitudes, however, remain largely unchanged, consistent with the trends identified earlier for the unconditioned quantities [see Fig.~\ref{fig:PiD_decomp_ions}) and associated discussion].
This behavior arises from the dependence of the conditioning on the current density. Since $J_{ez} \sim B_{in-plane}/\delta$, where $B_{in-plane}$ is the in-plane magnetic field strength and $\delta$ is the scale across which it changes, $\delta \sim \rho_e$, which becomes thinner as $T_e$ decreases, producing the seemingly counterintuitive effect. 
This effect is absent when $T_i$ is decreased while keeping $T_e$ fixed, because the structure of $J_{ez}$ remains controlled by electron physics and therefore retains its relevant scale. However, as discussed earlier in the context of unconditioned data, the amplitude does decrease, which is easily noticeable for shear [panels (c) to (e)], normal [panels (h) to (j)], and total [panels (m) to (o)] deformations. 

The strong point-wise anti-correlation between the ion shear and the normal deformations persists down to electron-scale current structures, indicating that localized ion phase space densities at these sub-ion scales are non-Maxwellian. The conditioning further highlights that even for ions, shear deformation largely accounts for most of the deformation, which was not easily visible in Fig.~\ref{fig:PiD_decomp_ions}. There are some exceptions, \textit{e.g.,} the current layer that is approximately 5 $d_i$ long at $y=18$ for the $T_e/T_i=0.25$ case, has $\mathrm{Pi-D}_i \approx {\rm Pi-D}_{\mathrm{normal},i}$.

\begin{figure}
    \centering
    \includegraphics[width=0.8\linewidth]{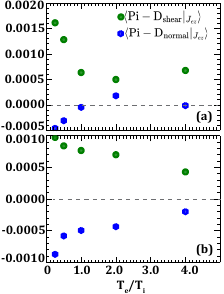}
    \caption{Net contribution by shear deformation  [denoted by green circles] $\langle {\rm Pi-D}_{\rm shear}|_{J_{ez}}\rangle$ and normal deformation [denoted by blue hexagons] $\langle {\rm Pi-D}_{\rm normal}|_{J_{ez}}\rangle$ in strong current sheet regions to changing internal energy of electrons [panel (a)] and ions [panel (b)].}
    \label{fig:heatingrate_conditioned}
\end{figure}

\section{Conclusions and Discussion \label{sec:conclusions}}
In the present work, using fully kinetic simulations of decaying turbulence, we analyze the effect of unequal electron-ion plasma beta, with total plasma $\beta \sim 1$, on the pressure-strain interaction. 
We focus on the spatiotemporally local variation in the total deformation part of the pressure-strain interaction, given by $\mathrm{Pi-D}_\sigma$.  We find that to better understand the variation in $\mathrm{Pi-D}_\sigma$, a deeper analysis of the decomposition of total deformation into shear and normal deformations is needed. 
Recent work on magnetic reconnection \cite{Adhikari_2025_PoP} has shown that in the presence of a strong mean (guide) magnetic field, $\mathrm{Pi-D}_\sigma \approx \mathrm{Pi-D}_{{\rm shear},\sigma}$ in the extended current sheet structures. However, one needs to be cautious of neglecting the normal deformation term. We find that, though small, the inclusion of $\mathrm{Pi-D}_{{\rm normal},\sigma}$ was still necessary in answering the trends seen in $\mathrm{Pi-D}_\sigma$ as a function of the initial plasma temperature ratio.  
Previous studies tested and found that the net change in the species' internal energy is well captured by the species' pressure-strain interaction \cite{Yang_2022_ApJ,yang_2024_JGR}, and other studies have used pressure-strain interaction as a proxy for the species' heating rate \cite{Roy_2022_ApJ} where it has been shown that the total deformation effects capture the heating rate, or rather the MHD-scale transfer rate. 

We also find that the net change in internal energy of both electrons and ions is positive, as expected, and it is due to the pressure-strain interaction that converts plasma bulk flow energy into plasma internal energy \cite{Yang17,yang_PRE_2017}. 
Ion heating, \textit{i.e.,} increase in ion internal energy, was found to dominate over electron heating, \textit{i.e.,} increase in electron internal energy, in previous studies that used similar simulation parameters \cite{parashar_2018_ApJL}, but had $T_e/T_i=1$ instead, and a similar observation was made in our simulation with $T_e=T_i$. However, in our simulations with $T_e>T_i$, we find that electron heating was slightly higher than ion heating, and similarly, we find that ion heating was slightly higher than electron heating in simulations with $T_i>T_e$, indicating that species with higher internal energy to begin with also gained slightly more internal energy. This could have important implications for magnetosheath \cite{Wang_2012_JGR} where $T_i \sim 5-10T_e$, for solar corona \cite{boldyrev_2020_PNAS} with $T_i \sim 4T_e$ and solar wind at 1 astronomical unit \cite{Wilson2018_ApJ} where $T_e \approx 1.6T_i$.
We also find that most of the conversion is carried by the deformation term, for both species, irrespective of the $T_e/T_i$ case,  as indicated by time cumulative sum of system-averaged ${\rm Pi-D}_\sigma$ following and matching the evolution of change in internal energy $\Delta \langle \mathcal{E}_{int,\sigma} \rangle$ [demonstrated in Fig.~\ref{fig:Eint_PS_PiD_PiDS_compare}]. Within %incompressible 
deformation-driven conversion of bulk to internal energy, we find that shear deformation plays a larger role, accounting for most of the change in the plasma internal energy (for both species). However, we discover that the contribution to the total deformation from only the shear part overestimates the net change in the species' internal energy, and this trend is seen to subside as the species temperature $T_\sigma$ is decreased [see Fig.~\ref{fig:Eint_PS_PiD_PiDS_compare}, green dotted lines], and that the contribution from the normal deformation is quite small. 
As the electron temperature decreases, both shear and normal flow deformation parts decrease, thus $\int^t_0 \langle {\rm Pi-D}_e \rangle dt$ decreases [panels (c) to (a), red dotted line] and so does $\int^t_0 \langle {\rm Pi-D}_{{\rm shear},e} \rangle dt$  [panels (c) to (a), green dotted line]. But normal flow gradients and diagonal pressure tensor elements decrease faster thus decreasing $\int^t_0 \langle {\rm Pi-D}_{{\rm normal},e} \rangle dt$ (not shown) faster, thereby improving the agreement between $\int^t_0 \langle {\rm Pi-D}_e \rangle dt$ and $\int^t_0 \langle {\rm Pi-D}_{{\rm shear},e} \rangle dt$ as $T_e$ decreases, while remaining largely unchanged when $T_e$ is kept constant [panels (c) to (e)]. 
However, in the case of the ions, as $T_i$ decreases, both shear and normal deformation parts %of the incompressible total deformation 
decrease, like electrons, but since ${\rm Pi-D}_{{\rm shear},i}$ and ${\rm Pi-D}_{{\rm normal},i}$ show strong local anti-correlation, unlike electrons, the decrease in ion normal and shear deformation parts causes a net decrease in ${\rm Pi-D}_i$ [see red dotted lines in Fig.~\ref{fig:Eint_PS_PiD_PiDS_compare}, panels (h) to (i)], alongside a decrease in only the shear component [see green dotted lines in Fig.~\ref{fig:Eint_PS_PiD_PiDS_compare}, panels (h) to (i)].

Our results show that electron shear deformation, ${\rm Pi-D}_{{\rm shear},e}$, dominates the %incompressible 
total volume-preserving deformation portion of pressure–strain interaction across all %$\beta_e$ 
cases and is concentrated in thin, electron-scale intermittent structures embedded between ion-scale eddies. Its amplitude decreases systematically as the initial electron temperature is reduced, primarily because the off-diagonal pressure tensor elements weaken more rapidly than the modest increase in flow-gradient scales that accompany lower temperatures. Although localized departures occur due to the time-dependent nature of decaying turbulence, the overall trend is robust. Normal deformation, ${\rm Pi-D}_{{\rm normal},e}$, is largely anti-correlated with the shear deformation but has a smaller amplitude, so including it slightly reduces the net ${\rm Pi-D}_e$ in regions where both coexist, underscoring the importance of retaining the normal term despite its smaller magnitude.
Similar to electrons, both ion shear and normal deformation quantities decrease in amplitude as the ion temperature is lowered, with occasional localized exceptions attributable to transient coherent structures.
However, the total deformation in ions is weaker than in electrons by nearly an order of magnitude. This follows from the smaller ion flow-gradient tensor, reflecting both the larger ion inertial scales and the slower ion flows. Moreover, for ions, the shear and normal deformation components are comparable in magnitude and strongly anti-correlated within ion-scale current sheets, and inside larger ion-scale structures, such as eddies, where demagnetized ions generate anisotropic pressure tensors, which thus suppresses the net local deformation ${\rm Pi-D}_i$.

To isolate how %strong 
structures at kinetic scales contribute to species internal energy density increase via the deformation term that is found in the strong current regions, we conditioned ${\rm Pi-D}_\sigma$ and its shear and normal components on strong current density, using $J_{ez}$ as the selection metric. Thresholds based on the standard deviation of the $J_{ez}$ distribution effectively captured the electron-scale current sheets that lie between large-scale eddies while excluding the core of the islands. Applying this mask shows that the strongest ${\rm Pi-D}_{\mathrm{shear},e}$ originates almost entirely from these intense current layers, and is anti-correlated at the electron scales with non-zero ${\rm Pi-D}_{\mathrm{normal},e}$.
Within the conditioned structures, shear deformation is predominantly positive and normal deformation is weakly negative, so the latter slightly reduces the net ${\rm Pi-D}_e$. This behavior reflects electron demagnetization in $d_e$-scale layers, the resulting enhancement of off-diagonal tensor elements ($P_{xz},P_{yz}$), and the comparatively weak in-plane gradients that drive normal deformation.
Conditioned quantities retain the same overall temperature dependence observed in the unconditioned data. Decreasing $T_e$ reduces the amplitudes of both electron shear and normal deformation through the associated weakening of diagonal pressure elements. Localized exceptions persist due to transient current-sheet dynamics as turbulence decays. 
For ions, the conditioning primarily reduces the spatial extent, rather than the strength, of the shear, normal, and total deformation when $T_e$ is lowered because the width of the electron-dominated current layers decreases with $\rho_e$, reducing the conditioning window. No analogous effect appears when $T_i$ is varied at fixed $T_e$, since the out-of-plane current is electron-dominated. % and the current-layer structure is electron-controlled.
Even when restricted to strong current sheets, ions exhibit a persistent local anticorrelation between shear and normal deformation, implying non-Maxwellian ion phase space densities exist down at electron scales. The conditioning also clarifies that ion shear deformation contributes most of the ${\rm Pi-D}_i$, similar to electrons, though isolated structures can temporarily exhibit ${\rm Pi-D}_i \approx {\rm Pi-D}_{\mathrm{normal},i}$.
Overall, the conditioned analysis shows that, at the electron-scale current sheets, shear deformation outpaces normal deformation, and thus contributes more to the bulk-to-internal energy conversion due to the net deformation. 

We note that the non-zero conditioned deformation does not mean a local increase or decrease in internal energy density of the species, as other terms from Eq.~\ref{eq:intengevolve} could play a role in local change of $\mathcal{E}_{int,\sigma}$. To test whether strong current sheet regions are the coherent structures where most change in plasma internal energy happens, we take the system average of the shear denoted as $\langle {\rm Pi-D}_{\rm shear}|_{J_{ez}}\rangle$ and normal, denoted by $\langle {\rm Pi-D}_{\rm normal}|_{J_{ez}}\rangle$, deformation components of ${\rm Pi-D}$, and analyze it as a function of $T_e/T_i$ at the time of analysis $t=30$. 
The results are shown in Fig.~\ref{fig:heatingrate_conditioned}.

Relative to the $T_e/T_i=1$ case, we find that as $T_i$ is decreased, the contribution to electron heating due to deformation at the strong current sheet regions remains somewhat the same. However, by decomposing into shear (green circles) and normal (blue hexagons) deformations [see panel (a) in Fig.~\ref{fig:heatingrate_conditioned}], we find the underlying cause. 
The shear deformation part remains quasi-steady around 0.0006, while the normal deformation part remains quasi-steady very close to zero, with very small variation in both.
%The shear part first dips slightly and then increases, whereas the normal part changes sign, becoming positive and then decreases for the $T_e/T_i=4$ case, because \textcolor{red}{why? what's the key physics here?}
For the same cases, unsurprisingly, ion $\langle {\rm Pi-D}_{\rm shear}|_{J_{ez}}\rangle$ and $\langle {\rm Pi-D}_{\rm normal}|_{J_{ez}}\rangle$ decrease in amplitude, thus making the former less positive and the latter less negative, which collectively produce the trend of decreasing net deformation-driven ion heating rate at the strong current sheets.
Looking at the case when $T_e$ is decreased instead, we find a sharp increase in electron $\langle{\rm Pi-D}_{\rm shear}|_{J_{ez}}\rangle$, suggesting, though, the amplitude and regions where shear deformation decreased as $T_e$ decreases [Fig.~\ref{fig:PiD_decomp_electrons_conditioned}], there are more regions where shear effects will locally increase electron internal energy than decrease it. As normal deformation is weakly anti-correlated at $d_e$-scale current sheets, the increase in electron $\langle{\rm Pi-D}_{\rm normal}|_{J_{ez}}\rangle$ means it becomes more negative, but not as fast as the shear deformation, thereby producing a net positive deformation-driven change to electron internal energy.
Due to previously discussed tightening of the current sheet regions as electron-scale current sheets thinned with $T_e$ decreasing, ion $\langle{\rm Pi-D}_{\rm shear}|_{J_{ez}}\rangle$ shows a trend of becoming more positive, whereas ion $\langle{\rm Pi-D}_{\rm normal}|_{J_{ez}}\rangle$ becomes more negative.
These trends answer why ion heating in strong current sheet regions happened ``around" the sheets, and not ``in" the sheets, as found in previous studies \cite{Parashar_2016_ApJ,Matthaeus_2016_ApJL,Servidio_2012_PRL,Dmitruk_2004_ApJ}, since the net effect of the discussed trends is that deformation-driven change to ion internal energy is very weak inside the electron-scale current sheet structures for all $T_e/T_i$ cases.

%\textcolor{red}{Discuss the big picture consequences of the results - how does it affect turbulent magnetosheath MMS measures, where $\beta \sim 1$. What about the solar wind? What about other astrophysical turbulent plasma systems?}

In light of these findings, there are several implications for turbulent space and astrophysical plasmas with $\beta \sim 1$. 
Our study sheds light on the role of shear vs. normal deformation in ``heating" plasmas when plasma species have different temperatures. The tendency of the initially hotter species to gain slightly more internal energy indicates that deformation-driven turbulent energization naturally helps maintain or modestly enhance the observed electron–ion temperature imbalance, rather than erasing it. Special care, however, is needed when analyzing the isotropic compression, \textit{i.e.}, pressure dilatation, part of pressure-strain interaction, which we did not study in the present work, and is found to be locally stronger than ${\rm Pi-D}_\sigma$ in observations \cite{bandyopadhyay_energy_2021}. The present work is a first step in understanding the interrelationship between sheared flow and particle ``heating" with applications to solar wind and other plasmas where Kelvin-Helmholtz-like vortex rollover exists, \textit{e.g.,} magnetopause flanks \cite{Nakamura2022PoP}, magnetopause surface/boundary layers \cite{Blasl2022_PoP}, and extended vortices in the magnetotail \cite{Turkakin2013JGR}.

A plethora of avenues for future work exist that could extend the results presented in this work. There are many computational challenges around simulating high-resolution decaying turbulence simulations,\textit{ i.e.,} high particle-per-grid, with well resolved grid-length, that lessen ``numerical heating" in simulations.  We focused on total plasma $\beta$ ranging from 1.5 to 2.4, (so about order 1,) but there are physical systems, such as solar wind ($\beta \sim 1$), Earth's magnetosheath ($\beta \sim 5-10$), Mars' magnetosheath ($\beta \sim 6-30$) and solar corona ($\beta  \ll 1$) that shows the spread in plasma beta.
Future kinetic PIC simulations that include cases reaching the Kawazura/Howes \cite{Kawazura_2019_PNAS,Howes_2024_JPP} regime boundaries can test whether the shear and normal deformation trends found in the present study and localization trends persist, with and without conditioning on parameters, such as strong current sheet regions.
The present analysis uses undriven, decaying turbulence, but there exist natural plasma systems where turbulence is instead driven, such as solar wind, solar corona, and astrophysical jets, where continuous energy input comes from velocity shears \cite{Kintner_1976_JGR_Shear_observations,Gentle_2010_PST_shear_lab,Goodwill_2025_PoP,Antoni1999_Shear}, compressive effects \cite{Kawazura_2020_PRX,Zhdankin_2021_ApJ,Federrath_2010_compressive,Makwana_2020_compressive}, magnetic field gradients \cite{Toufen2022gradient_PoP}, reflection-driven \cite{Meyrand_2025_JPP}, pressure-anisotropy-driven \cite{Melville_2016_MNRAS}, electron-temperature-gradient-driven \cite{Adkins_2022_JPP,Chapman2025_PRR} among other myriad of driving mechanisms.
Furthermore, current simulations are 2.5D, with no variations allowed in the out-of-plane direction. In fully 3D turbulence simulations that are also driven, intermittency and reconnection statistics can be different, thus altering where and how ${\rm Pi-D}_\sigma$ is produced. 
Future work should focus on performing driven 3D fully kinetic simulations with varying $T_e/T_i$ and check whether shear deformation remains the dominant effect contributing to deformation in strong structures for both species in strong current sheet regions, and the cumulative contribution to changing species' internal energy from the whole simulation domain vs. only the strong current sheet regions.

We conclude our discussion with two caveats of the present work. Firstly, since pressure-strain interaction and the total deformation (contained within ${\rm Pi-D}_\sigma$) and the shear and normal deformation parts are intimately connected to the fluid-like description of energy evolution, they are mechanism-agnostic. In recent years, a significant effort has led to identifying physical mechanisms in turbulent systems that drive the local change in energy spatiotemporally locally, by looking at the phase-space signatures \cite{Klein16}. By doing so, contributions to the net change in energy have been linked back to the disparate populations routinely found in non-LTE turbulent plasmas \cite{Servidio_2017_PRL,larosa2025velocity}, thus quantifying each population's contribution \cite{Klein16,Klein17,Klein20,Chen19,TCLi_2019_JPP,Afshari_2021_JGR}. A similar outlook into pressure-strain interaction has only been recently discovered \cite{Conley_2024_PoP,Barbhuiya_2026_KPS}, and further work is needed to isolate how different mechanisms lead to shear and normal deformations for fundamentally non-LTE phase space densities, and how their phase-space signatures differ, and how different populations contribute to deformation physics.
Secondly, for fundamentally non-LTE plasmas, change to plasma energy only captures changes to the low-order moments of the phase space densities, and only recently has it been shown for weakly collisional plasmas that the changes to the higher order moments of the phase space densities can be equally, if not more, important \cite{Cassak_FirstLaw_2023,Barbhuiya_PRE_2024}, locally and globally. The present work did not attempt to quantify how higher-order non-equilibrium terms captured by the so-called HORNET effective power density \cite{Barbhuiya_PRE_2024} change as a function of unequal electron-ion plasma beta in weakly collisional plasma turbulence.

\begin{acknowledgments}
%\textcolor{red}{Unfunded work, however, simulations were performed on NERSC, so we will need to acknowledge that, and thus acknowledge the repo award that Paul has. Still using the company laptop, so will need to acknowledge the grants that paid for it.}
M.H.B gratefully acknowledges support from NASA Grant 80NSSC24K0172 and 80NSSC23K0409, NSF Grant PHY-2308669, and DOE Grant DE-SC0020294. 
S.A. is supported by Plasma Physics program NSF Division of Physics grant PHY-2108834.
This research used resources of the National Energy Research Scientific Computing Center (NERSC), a U.S. Department of Energy Office of Science User Facility located at Lawrence Berkeley National Laboratory, operated under Contract No. DE-AC02-05CH11231 using NERSC award FES-ERCAP0027083, which was awarded to Dr. Paul Cassak.
\end{acknowledgments}

\section*{Data Availability Statement}
The data supporting the findings of this study are openly available in Zenodo at 
https://doi.org/10.5281/zenodo.17885575

%used for making the simulation data plots in the figures will be uploaded to Zenodo, alongside how to use it to replicate the plots.

%\appendix

%\section{Appendixes}

%\nocite{*}
\bibliography{Thermal_Disequilibration}% Produces the bibliography via BibTeX.

\end{document}